# Interlayer excitons in semiconductor bilayers under a strong electric field

S. Kovalchuk[1], K. Greben[1], A. Kumar[1], S. Pessel[1], K. Watanabe[3], T. Taniguchi[3], D. Christiansen[2], M. Selig[2], A. Knorr[2], K.I. Bolotin[1*]

[1] Physics Department, Free University of Berlin, Germany
[2] Physics Department, Technical University of Berlin, Germany
[3] National Institute for Materials Science, Tsukuba, Japan

**Abstract**

Excitons in bilayer transition metal dichalcogenides (2L-TMDs) are Coulomb-bound electron/hole pairs that can be viewed as broadly tunable analogs of atomic or molecular systems. Here, we study the properties of 2L-TMD excitons under strong electric field. To overcome the field limit, reached in previous experiments, we developed a new organic/inorganic molecular gating technique. Our approach allows reaching the field > 0.27 V nm$^{-1}$, about twice higher than previously available. Under this field inter and intra-layer excitonic are brought into an energetic resonance, allowing us to discover new emergent properties of the resulting hybridized states. First, as the result of hybridization, intralayer excitons acquire an interlayer character. Second, the same hybridization allows us to detect new excitonic species. Third, we observe an ultra-strong Stark splitting of > 380 meV with exciton energies tunable over a large range of the optical spectrum, with potential implications for optoelectronics. Our work creates new possibilities for using strong electric fields to unlock new physical regimes and control exciton hybridization in 2D heterostructures and other systems.

**Introduction**

Interlayer excitons (IX) in bilayer transition metal dichalcogenides (2L-TMDs) are Coulomb-bound pairs of electrons and holes localized to different layers of these two-dimensional semiconductors. Compared to intralayer excitons, the electron/hole separation in IXs is larger and the oscillator strength lower[1–7]. As a result, IXs feature lifetimes in the tens of nanoseconds[2,8,9] and diffusion lengths up to microns[10,11], much higher than that of intralayer excitons. These properties led to an explosion of interest in IXs in fields such as excitonic transport[10,12], Bose-Einstein condensation[13,14], excitonic insulators[15,16], and quantum simulation[17]. Another distinguishing property of IXs is their coupling to intralayer excitons resulting from interlayer hole tunneling[4,5,7]. This mechanism leads to a tunable enhancement of the oscillator strength of IXs in some 2L-TMDs, e.g. 2L-MoS$_2$, allowing their observation via absorption spectroscopy[4,6,7]. Finally, IXs exhibit a strong Stark splitting in an electric field oriented perpendicular to the plane of the material due to the large static dipole moment of IXs[2,4,6,7,18,19]. Because of that, the energy position, oscillator strength and coupling strength to other excitonic species can be tuned by the electric field.

The perpendicular electric field in 2L-TMDs is conventionally applied in a dual-gated field effect transistor geometry. Electrostatic gates consisting of dielectric (e.g. hBN or SiO$_2$) and conductor (e.g. gold, Si or graphene) are assembled on both sides of the 2L-TMD. In such a configuration, the difference between gate voltages applied to the top and bottom conductors controls the field across the material, while the sum of gate voltages controls the carrier density and Fermi level[20]. Generally, the strength of the perpendicular field controls the energy splitting between IXs with oppositely oriented dipole moments (denoted IX$_+$ and IX$_-$). The maximum reported splitting in conventional dual-gated devices, 200 meV[7,21–23], is in practice limited by the breakdown of the dielectric material. At this point of dielectric breakdown, the field inside the 2L-TMD reaches ≈ 0.15 V nm$^{-1}$. The IX splitting under that field is comparable to the spin-orbit splitting of the 2L-TMD and in general is not sufficient to bring IXs in and out of resonance with other excitonic species, such as intralayer excitons. Hence, until now, investigations were restricted to low electric fields, setting an overall limit for studying the coupling of different exciton species. While an order of magnitude higher electric field has been recently generated using ionic liquids[20,24], that approach is so far limited to room temperature and makes optical measurements challenging.

Here, to study the regime of a tunable coupling between IXs and other excitonic species, we overcome the limits of previous gating technologies. We develop a hybrid molecular gating approach allowing the generation of an electric field of > 0.27 V nm$^{-1}$ (displacement field > 2 V nm$^{-1}$), doubling the previous limit. We demonstrate tuning of the splitting energy of IXs > 380 meV. In this high-field regime, we find signatures of new types of excitons that have not been seen before: a high-energy interlayer exciton IX$_2$ and a dark interlayer exciton.

**Device concept.** We study the effect of a strong electric field in a 2L-TMD on top of a few-layer hBN. The thin layer of hBN serves to suppress substrate scattering. We start with 2L-MoS$_2$ because of its well-characterized bright interlayer excitons[3] and then turn to 2L-MoSe$_2$. To overcome the limits of conventional gating, we add layers of charges next to the heterostructure (Fig. 1a). The top layer consists of F$_4$TCNQ molecules, a well-known molecular acceptor[25–31] (charge density $\sigma_t$). The bottom layer originates from the donor states already present at the interfaces between the TMD and the SiO$_2$/Si stack. These states, with charge density $\sigma_b$, arise due to a combination of photodoping[32–34] and interface charge trapping[35,36]. A high electric field inside the 2L-TMD is generated due to high densities of top and bottom charges. Furthermore, we can adjust that field by varying the voltage V$_G$ between Si and the 2L-TMD. In order to calculate the electric field in the 2L-TMD, we model our system as three capacitors connected in series (Fig. 1a), with the middle one corresponding to the layers of the 2L-TMD. The field inside the 2L-TMD is given by $F_Z = \frac{eC_{BL}}{\varepsilon_0 \varepsilon_{TMD}} \left( E_F^{top} - E_F^{bot} \right)$, where $E_F^{top}$ and $E_F^{bot}$ are the Fermi energies of the top and bottom TMD layers, $C_{BL}$ is the areal capacitance of the bilayer, and $\varepsilon_{TMD}$ is its dielectric constant. Assuming that the Fermi level of the 2L-TMD lies within the bandgap of each layer and the screening due to free carriers is absent, we estimate the field inside the 2L-TMD (SI note S1):

$$F_Z \approx \frac{1}{2\varepsilon_{TMD}} (\sigma_t - \sigma_b - eV_G C_G), \quad (1)$$

where $C_G$ is the areal capacitance between the bilayer and the Si gate. The formula shows that the field inside the 2L-TMD can exceed the field inside the SiO$_2$, the last term inside the brackets. It is important to note that in addition to the field strength, the gate voltage $V_G$ controls the Fermi level inside the 2L-TMD, which, in turn, defines the charge density $\sigma_t$ (Fig. 1a). The field is maximized when the Fermi level reaches the bottom of the conduction band. At this point, the screening due to free carriers induced in the 2L-TMD prevents an additional field increase.

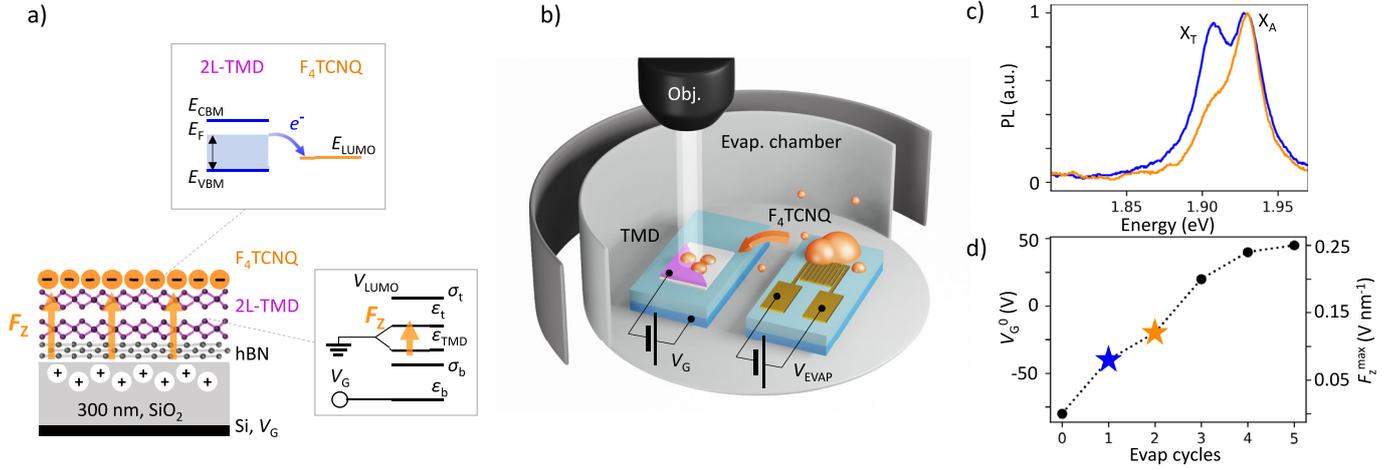

***Figure 1: Hybrid molecular gating. (a)*** *Device schematic of a 2L-TMD with two layers of charge, and an equivalent 3-capacitor circuit. The top inset illustrates the energetic alignment between 2L-MoS$_2$ and F$_4$TCNQ.* ***(b)*** *Evaporation chamber setup schematic. Organic molecules are deposited onto a TMD in situ by controllably heating the coil on a separate chip.* ***(c)*** *Photoluminescence spectra of 2L-TMD before (blue) and after (orange) evaporation of F$_4$TCNQ, at V$_G$ = -20 V. The decrease of the trion peak intensity indicates that 2L-MoS$_2$ becomes more neutral due to hole doping via charge transfer.* ***(d)*** *The smallest voltage at which the trion feature is visible in the optical spectra (V$_G^0$) vs. the number of the evaporation cycle. The electric field extracted for that molecular coverage is shown on the right axis. Molecular density at points shown as blue and orange symbols correspond to the same color curves in (c).*

We develop a new technique of *in situ* evaporation to control the molecular density $\sigma_t$ – and with it $F_Z^{max}$ – *in situ* during the experiment. It works by applying a short pulse of current to the metal coil microfabricated on a separate chip, evaporating molecules. The chip is placed close to the measured sample inside the cryostat (Fig. 1b, details in *Methods)*. This approach adds more flexibility compared to traditional deposition techniques[26,37–39], and allows for precise control of the surface coverage, which could be even more crucial when using other organic molecules, e.g. dyes[40].

To determine the charge densities of the top and the bottom charge layers, $\sigma_t$ and $\sigma_b$, we use optical spectroscopies in the setup shown in Fig. 1b. We measure photoluminescence (PL) and reflectivity spectra in the region of the intralayer neutral exciton (X$_A$ at 1.94 eV) and intralayer trion (X$_T$ at 1.91 eV) of the 2L-MoS$_2$[26]. When the Fermi level of the TMD is inside the bandgap, only the neutral excitons are seen (Fig. 1c, orange) [26,41]. When it reaches the conduction band minimum, neutral excitons bind to the resident free carriers forming trions[39,42,43] (Fig. 1c, blue). The minimum voltage $V_G^0$ when this starts to occur is directly related to $\sigma_t$ and $\sigma_b$ (SI note S2). From these values, we estimate the highest field via Eq. 1. For the highest molecular coverage, we get $F_z \approx 0.25$ V nm$^{-1}$ (Fig. 1d), significantly higher than the limit of conventional dielectrics.

**Stark splitting in bilayer TMD systems.** We now study the effect of the strong electric field on interlayer excitons in 2L-MoS$_2$. Figure 2a shows the differential reflectivity map of a molecularly gated 2L-MoS$_2$ vs. $V_G$. We identify the spectral features corresponding to intralayer X$_A$ and X$_B$ excitons (1.94 and 2.11 eV at $V_G$ = -80 V) as well as interlayer exciton IX$_1$ (2 eV at $V_G$ = -80 V) that undergo the Stark splitting in a non-zero electric field (Fig. 2c). To understand the field dependence of these excitons, we find the solution for this electrostatic system using our capacitor model (Fig. 1b, details in SI note S1). For every $V_G$ value, we find charge densities and Fermi energies of top and bottom TMD layers (Fig. 2b). We then determine the field strength from Eq. 1 and the expected Stark shift of interlayer excitons $E_{IX1\pm} = E_{IX1}^0 \pm F_Z \times d_{BL}$. Here $E_{IX1}^0 = 2$ eV is the spectral position of IX$_1$ at zero field (denoted IX$^0_1$ in Fig.

2a) and $d_{BL}$ = 0.6 e·nm is its dipole moment, very close to the TMD interlayer distance[1]. The calculated positions of the two split IXs as a function of $V_G$ are shown as dashed lines in Fig. 2a. The right axis shows the electric field corresponding to the gate voltage. We see that the observed spectral shifts of the IX$_{1\pm}$ excitons match our simple model in the limit of small $F_Z$ ($V_G$ from -80 V to -40 V). When the field becomes large enough to bring inter- and intra- layer excitons close to an energetic resonance, we see a more complex avoided-crossing pattern and excitonic shifts become non-linear. Such avoided crossing has been previously shown in 2L-MoS$_2$[4,6,7,44].

At $V_G$ = -80 V, two interlayer components IX$_{1+}$ and IX$_{1-}$ are doubly degenerate at $F_Z$ = 0 (Fig. 2c, bottom). The Fermi level of the TMD is equal to $E_{LUMO}$ and $\sigma_t$ = 0; the fields from $\sigma_b$ and gate electrodes are compensated. When $V_G$ = 0, there is no field from the back gate, however we see a splitting of 200 meV between IX$_{1+}$ and IX$_{1-}$, corresponding to $F_Z$ = 0.16 V nm$^{-1}$. This field is generated only by the two planes of charge $\sigma_t$, $\sigma_b$ (Fig. 2c, middle). Finally, for positive V$_G$, this field increases until $F_Z$ = 0.24 V nm$^{-1}$, and the corresponding splitting between IX$_{1+}$ and IX$_{1-}$ reaches 300 meV. At this point, the Fermi level of the 2L-TMD reaches the conduction band minima and the field does not increase further as free carriers begin to screen it (Fig. 2c, top).

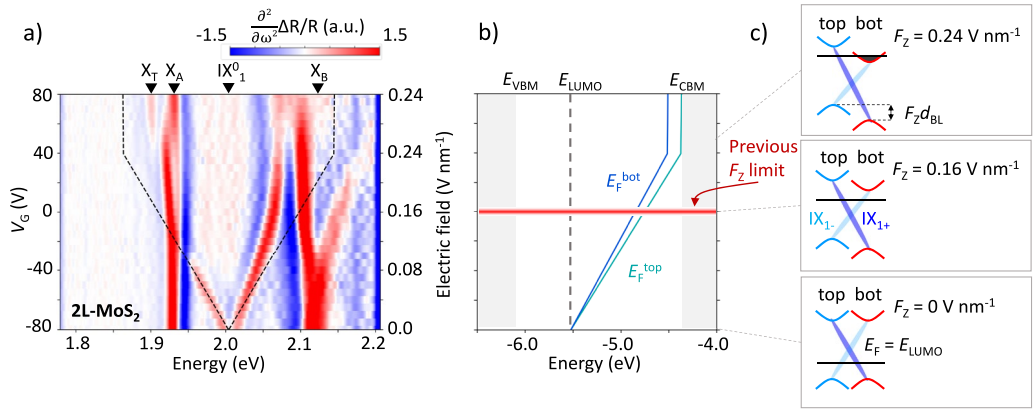

*Figure 2: Measuring excitonic response under strong electric field. (a) Map of the second derivative of reflectivity contrast ($R_C = \Delta R/R$) vs. back gate voltage (left axis) and the electric field extracted from the electrostatic model (right axis) in 2L-MoS$_2$. The field increase stops at 0.24 V nm$^{-1}$ due to screening from free carriers. The features corresponding to intralayer ($X_A$, $X_B$, $X_T$) and interlayer (IX$_1$) excitons at $V_G$ = -80 V are marked. Dashed lines indicate the Stark splitting of IX$_1$ in the simple model that does not account for inter-excitonic interactions. (b) The Fermi energies of the top and bottom layers in the 2L-TMD were obtained by solving the capacitor model visualized in Fig. 1a. The LUMO level of F$_4$TCNQ is shown as a gray line; the field strength limit of conventional dielectric technologies is shown in red. (c) Sketches showing the composition of interlayer excitons for different field strengths. IX$_{1+}$ is shown as dark-blue, and IX$_{1-}$ as a light-blue. The black line indicates the Fermi level.*

**Coupling between inter- and intralayer excitons.** We now focus on the field-dependent interactions between intra- and interlayer excitons. In general, it is known that the interlayer IX$_1$ exciton couples to the intralayer exciton X$_B$ via hole tunneling[1,4] (arrows in Fig. 3a; T denotes the tunneling strength parameter). This means that the state IX$_1$ partially acquires the intralayer character of X$_B$ and hence deviates from the linear field dependence[1,4,7,45]. Conversely, the state X$_B$ hybridizes with IX$_1$ acquiring an interlayer character and splits into two components X$_{B+}$ and X$_{B-}$ in the field. To describe this scenario quantitatively, we extract the energies of all excitonic peaks (Fig. 3b, diamonds) vs. field and fit them to a model based on the Bloch equations formalism[4] (solid lines, details in SI note S4). This model implies that IX$_{1-}$ couples to X$_{B-}$ only and IX$_{1+}$ to X$_{B+}$ only and neglects other more complex types of couplings[44,46]. The energies of X$_B$ and IX$_1$ at zero field unperturbed by interexcitonic interactions, dipole moment $d_{BL}$, as well the hole tunneling strength $T$ are free parameters of the model. We see that the model describes the observed positions and amplitudes well (SI fig. 6, 7). We extract the tunneling strength $T = 40.7 ± 0.4$ meV. While similar values have been reported previously[7,44], the larger electric field range accessible in our experiments results in a high accuracy of this estimate, and allow us to resolve the shift of the split $X_{B-}$ resonance due to coupling to $IX_{1-}$.

The model in Ref. 4 suggests that the intralayer exciton $X_A$ does not couple to $IX_1$ as the required tunneling is spin-forbidden (opposite color of bands in Fig. 3a). The $X_A$ energy is then expected to stay field-independent, other than the minimal (<1 meV) contribution from the quantum-confined Stark Effect[19]. In contrast, we see $X_A$ splitting into two peaks at fields higher than 0.15 V nm$^{-1}$ (red and orange lines in Fig. 3b, the data corresponding to extracted peaks is shown in Fig. 3c). The splitting reaches 15 meV at the highest field. To account for this phenomenon, we introduce another interlayer exciton, $IX_2$, into our model [45]. That exciton is composed of an electron wavefunction localized in the top conduction subband of one layer and a hole wavefunction in the bottom subband of another layer (Fig. 3a). The $IX_2$ exciton (that can also be labelled as intralayer B exciton) is expected to couple to $X_A$ via the same hole hopping mechanism that couples $IX_1$ to $X_B$ (Fig. 3a). From symmetry arguments, the coupling strength between $X_A$ and $IX_2$ is the same as between $X_B$ and $IX_1$ [45]. We also assume that $IX_2$ has the same dipole moment as $IX_1$. We now fit the position of $IX_2$ using unperturbed positions of $IX_2$ and $X_A$ as our only free parameters (red and orange solid lines in Fig. 3b). The model that includes coupling between $X_A$ and $IX_2$ describes the $V_G$-dependent energy of $X_A$ surprisingly well. We find $E_{IX2} = 2.168 ± 0.003$ eV, 164 meV above $IX_1$, at zero field. To the best of our knowledge, the large magnitude of the electric field in our experiment allows the first experimental observation of $IX_2$. In general, $IX_2$ cannot be observed through absorption spectroscopies due to its low oscillator strength stemming from the large separation from the $X_A$ exciton (the coupling to the $X_B$ exciton is the reason $IX_1$ can be observed directly).

One interesting aspect of our model is that it also gives the unperturbed energies of the excitons. This is in contrast to observed energies of excitons that are affected by interexcitonic interaction even at zero field. While the interaction of course cannot be turned off in a real material, it is strongly reduced in the 3R phase of MoS$_2$, where hole tunneling is symmetry forbidden[5]. The separation between $X_A$ and $X_B$ is reduced by 35 meV in the 3R phase compared to the 2H phase[5]. This is comparable, in our model, to the 27 meV change of the $X_A$ and $X_B$ energy difference with introduction of coupling to $IX_1$ and $IX_2$.

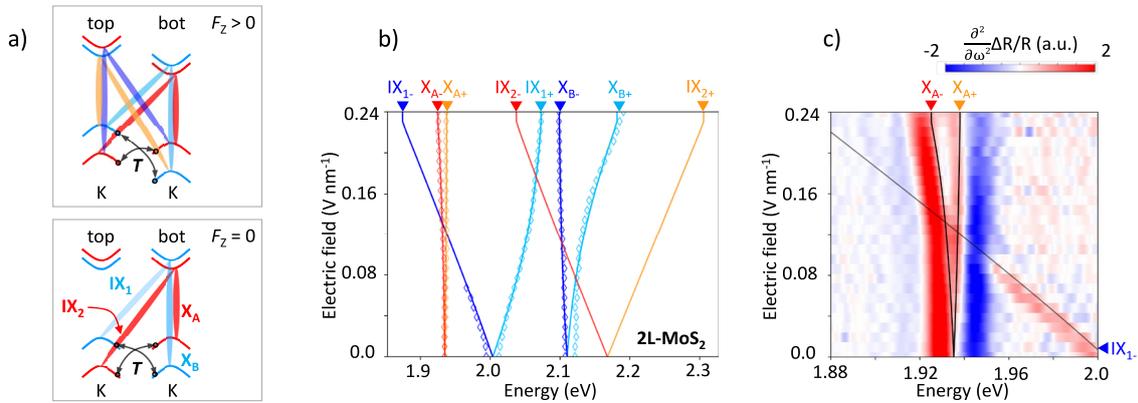

*Figure 3:* **Couplings between IXs and intralayer excitons.** *(a) The configurations of various interlayer and intralayer excitons in 2L-TMDs at zero and non-zero electric fields. The coupling between the excitons sharing electron wavefunctions is mediated by spin-conserving interlayer tunneling (arrows). (b) The second derivative of reflectivity contrast spectra, region of $IX_1$ crossing $X_A$. The y-axis corresponds to the*

-80 V to 45 V region in Fig. 2a. *(c)* The dependence of the excitonic peak energies extracted from the data of Fig. 2a on the electric field (diamonds) along with theoretical predictions based on the Bloch equations (lines). Note the new interlayer exciton $IX_2$, the coupling with which leads to the splitting in the energy of $X_A$ at high field.

We now turn to another bilayer material from the TMD family, MoSe$_2$. In Fig. 4a we plot the gate-dependent second derivative of the reflectivity contrast for that material. The first thing to note is the significantly larger electric field achievable in MoSe$_2$, 0.27 V nm$^{-1}$ (or the displacement field $D_{MOSE2}$ = 1.98 V nm$^{-1}$) compared to 0.24 V nm$^{-1}$ ($D_{MOS2}$ = 1.63 V nm$^{-1}$) in 2L-MoS$_2$. This increase in field originates from the smaller separation between the valence band of the semiconductor and the LUMO level of F$_4$TCNQ in the case of MoSe$_2$. This results in higher maximum charge transfer from that material. Using the same analysis as above, we characterize excitons in 2L-MoSe$_2$. We find the dipole moment of $d_{BL}$ = 0.65 e·nm for interlayer excitons and exciton tunneling strength $T$ = 44.6 ± 2.1 meV. The parameter $T$ is 10 meV lower compared to the calculation[45,47]. At our maximum field, $IX_{1-}$ and hybrid state $X_{B+}$ experience ultra-strong Stark splitting of > 380 meV. Finally, due to a larger energy difference between $X_A$ and $IX_2$ (369 meV in MoSe2, compared to 218 meV in MoS2) the splitting of $X_A$ is smaller and less pronounced (Fig. 4b), reaching 6 meV at the highest field.

A new feature present only in MoSe$_2$ is a weak signature of the avoided crossing between $X_A$ and a state 30meV below $IX_1$, visible in the reflectivity map (zoomed in around $X_A$ in fig. 4b, at $F_Z$ = 0.09 V nm$^{-1}$). That energy is consistent with the spin- or momentum- dark interlayer exciton $IX_d$ shown in Fig. 4d. We note that two interlayer states (dark $IX_d$ and $IX_1$) are expected to be almost degenerate in MoS$_2$, due to a much smaller conduction band splitting in that material (3 meV vs. 19 meV)[47–49]. We obtain additional information about this crossing from PL spectra in Fig. 4c. In addition to the $X_A$ exciton, we see a feature typically attributed to the momentum indirect K-Γ exciton at 1.45eV. The feature blueshifts with field, suggesting a partially interlayer character of that state. The observed out-of-plane dipole moment of the K-Γ exciton is 0.08 e·nm, smaller than in the previous reports[49]. Interestingly, we see that both $X_A$ and K-Γ states are brightened (by a factor of 4 for K-Γ and 1.2 for $X_A$, details in SI fig. 8) exactly at $F_Z$ = 0.09 V nm$^{-1}$, the field corresponding to the crossing. We hypothesize that this is related to the hybridization between $X_A$ and dark $IX_d$ when they are brought into resonance. The population of K-Γ excitons, sharing an electron with A states[47–49], increases correspondingly.

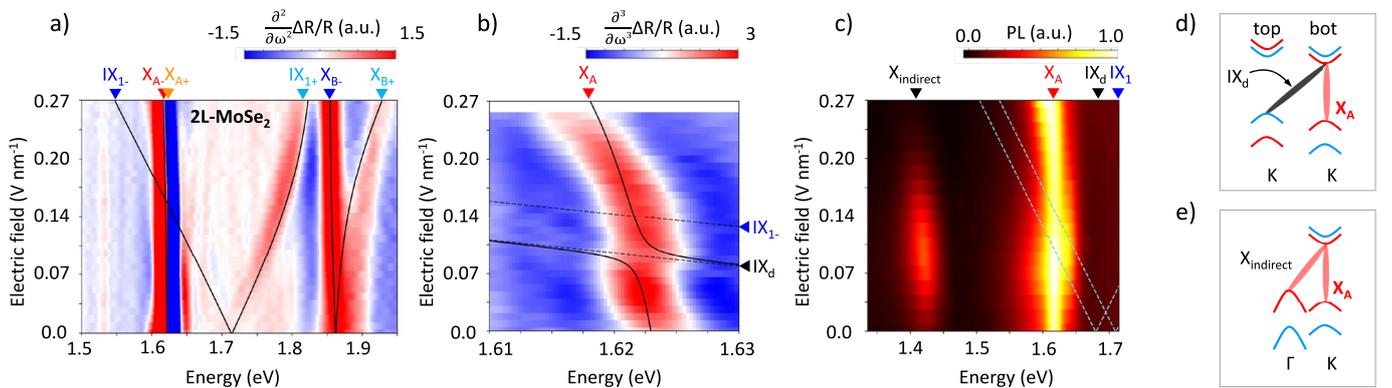

*Figure 4: 2L-MoSe$_2$ in a strong electric field.* *(a)* The second derivative of reflectivity contrast for 2L-MoSe$_2$. *(b)* The third derivative of reflectivity contrast, centered on the avoided crossing region. Dashed lines show the expected energies of the allowed $IX_1$ exciton, and spin dark interlayer exciton. Solid lines show an example avoided crossing between $IX_d$ and $X_A$, with coupling strength of 3 meV. *(c)* PL map with

*the lowest lying indirect state, and $X_A$ exciton.* **(d)** *Dark spin forbidden exciton schematic.* **(e)** *Indirect exciton schematic.*

## Discussion

The electric field achieved via hybrid molecular gating roughly doubles the limit achievable with dielectric gates. In fields up to 0.27 V nm$^{-1}$, we observe: 1) coupling between the excitons $IX_1$ and $X_B$, 2) signatures of avoided crossing in the intralayer exciton $X_A$ attributable to a new interlayer exciton $IX_2$, and 3) signatures of coupling between a dark interlayer exciton and $X_A$.

The maximum electric field can be increased further by using molecules with a lower LUMO level (e.g., $F_6$TCNQ with a LUMO level 0.3 eV lower than $F_4$TCNQ[37,38]). The charge transfer – and hence the field – can be further increased by using a strong donor molecule underneath the 2L-TMD layer. It is worth noting that across two example devices presented here we almost continuously tune the ensemble of absorption resonances in the range between 1.5 and 2.2 eV, a significant portion of the visible spectrum. It makes doubly molecularly gated 2L-TMDs attractive optoelectronic materials.

We anticipate several interesting uses of new excitonic states at high electric fields. First, the response of an exciton to the out-of-plane field indicates its out-of-plane character and can be used, in principle, for the "fingerprinting" of excitonic species. As we have shown, the field response of some states (e.g. $X_A$) is only resolvable in high enough fields. Therefore, the ability to apply high electric fields may enable a more detailed identification of various excitonic species with debated character. Second, the state $IX_2$ reported here is an attractive candidate to transmit information in excitonic circuits. That state is normally dark and should have a very long lifetime enabling its propagation over long distances. The information encoded in that state is "written" or "read out" in the regions of the circuit exposed to a high electric field. In those areas, the state is brought into an energetic resonance with an intralayer exciton thereby increasing its coupling to light. Finally, we expect our technique for generating strong electric fields to apply to many other 2D materials and heterostructures. New excitonic species and phases of matter should appear when the strength of the electric field is high enough. A particularly compelling example is the predicted emergence of an excitonic insulator phase when the exciton binding energy becomes smaller than the bandgap of a system exposed to a strong electric field[15,16,50–54].

## Methods

**Device fabrication.** Samples were prepared using a PDMS dry stamping method, and transferred onto hBN directly exfoliated onto a 300 nm $SiO_2$/Si chip[55]. Contacts were made using electron beam lithography (EBL) followed by thermal evaporation of Cr/Au (3 nm/70 nm). All samples were cleaned by AFM "nano-squeegee" (60 nN force) to clean the surface and improve the contact with molecules[56,57].

***In situ* evaporation.** In our technique, we place a small amount of organic acceptors $F_4$TCNQ (Sigma-Aldrich, amount < 1 mg) onto an evaporation coil fabricated on a 300 nm $SiO_2$/Si chip. The coil is made using EBL with the same parameters as the contacts on the sample. The coil resistance is 60 Ohm, and the design is similar to Ref. 41. This chip is loaded into our optical cryostat right next to the 2L-TMD. To evaporate a controlled dose of molecules, we apply a short voltage pulse to the evaporator coil (Fig.1b, $V_{EVAP}$, duration is selected between 1s and 3s), heating the molecules above their melting temperature. During the heating process the temperature of the 2L-TMD remains virtually unchanged (details in SI fig. 10). The evaporation chamber is sealed inside the inner heatshield of the cryostat to avoid contamination. This *in situ* evaporation approach has multiple advantages. First, the density of molecules can be adjusted during the experiment without heating the device. Second, evaporation at cryogenic temperatures solves the problem associated with molecules agglomerating into clusters which occurs during room-temperature deposition [26]. Finally, we avoid the exposure of a thin molecular layer to the ambient environment.

**Optical measurements.** We use a home-built confocal PL/reflectivity setup at cryogenic temperature (4K). PL measurements were done using a 532 nm continuous wave laser. Reflectivity measurements were carried out using a broadband pulsed supercontinuum laser source (SuperK, 400 – 1000 nm). All measurements employed SLWD Nikon objective (0.4 NA, 22mm WD) and an Andor spectrometer (300 and 600 lines/mm gratings). To extract the positions of excitonic resonances from Fig. 2a and 4a, we used Kramers–Kronig relation to model the dielectric function (see SI note S3 for details).

**Acknowledgements**

We acknowledge useful conversations with Nele Stetzuhn, Ben Weintrub, Denis Yagodkin, and Georgy Gordeev. We acknowledge the financial support from Deutsche Forschungsgemeinschaft (SfB951 and TRR227) and BMBF (05K2022 – ioARPES).

**References**

1. Deilmann, T. & Thygesen, K. S. Interlayer Excitons with Large Optical Amplitudes in Layered van der Waals Materials. *Nano Letters* **18**, 2984–2989 (2018).

2. Wang, Z., Chiu, Y. H., Honz, K., Mak, K. F. & Shan, J. Electrical Tuning of Interlayer Exciton Gases in WSe2 Bilayers. *Nano Letters* **18**, 137–143 (2018).

3. Gerber, I. C. *et al.* Interlayer excitons in bilayer MoS 2 with strong oscillator strength up to room temperature. *Phys. Rev. B* **99**, 035443 (2019).

4. Lorchat, E. *et al.* Excitons in Bilayer MoS 2 displaying a collosal electric eld splitting and tunable magnetic response. *Physical Review Letters* **126**, 37401 (2021).

5. Paradisanos, I. *et al.* Controlling interlayer excitons in MoS2 layers grown by chemical vapor deposition. *Nat Commun* **11**, 2391 (2020).

6. Peimyoo, N. *et al.* Electrical tuning of optically active interlayer excitons in bilayer MoS2. *Nature Nanotechnology* **16**, 888–893 (2021).

7. Leisgang, N. *et al.* Giant Stark splitting of an exciton in bilayer MoS2. *Nature Nanotechnology* **15**, 901–907 (2020).

8. Rivera, P. *et al.* Observation of long-lived interlayer excitons in monolayer MoSe2–WSe2 heterostructures. *Nat Commun* **6**, 6242 (2015).

9. Miller, B. *et al.* Long-Lived Direct and Indirect Interlayer Excitons in van der Waals Heterostructures. *Nano Lett.* **17**, 5229–5237 (2017).

10. Unuchek, D. *et al.* Room-temperature electrical control of exciton flux in a van der Waals heterostructure. *Nature* **560**, 340–344 (2018).

11. Jauregui, L. A. *et al.* Electrical control of interlayer exciton dynamics in atomically thin heterostructures. *Science* **366**, 870–875 (2019).

12. Yuan, L. *et al.* Twist-angle-dependent interlayer exciton diffusion in WS2–WSe2 heterobilayers. *Nat. Mater.* **19**, 617–623 (2020).

13. Wang, Z. *et al.* Evidence of high-temperature exciton condensation in two-dimensional atomic double layers. *Nature* **574**, 76–80 (2019).

14. Blatt, J. M., Böer, K. W. & Brandt, W. Bose-Einstein Condensation of Excitons. *Phys. Rev.* **126**, 1691–1692 (1962).

15. Gu, J. *et al.* Dipolar excitonic insulator in a moiré lattice. *Nature Physics* **18**, 395–400 (2022).

16. Xu, Y. *et al.* Tunable bilayer Hubbard model physics in twisted WSe2. (2022).

17. Montblanch, A. R.-P. *et al.* Confinement of long-lived interlayer excitons in WS2/WSe2 heterostructures. *Commun Phys* **4**, 1–8 (2021).

18. Zhao, Y. *et al.* Interlayer exciton complexes in bilayer MoS2. *Physical Review B* **105**, L041411 (2022).

19. Roch, J. G. *et al.* Quantum-Confined Stark Effect in a MoS2 Monolayer van der Waals Heterostructure. *Nano Letters* **18**, 1070–1074 (2018).

20. Weintrub, B.I., Hsieh, YL., Kovalchuk, S. *et al.* Generating intense electric fields in 2D materials by dual ionic gating. *Nat Commun* **13**, 6601 (2022).

21. Hattori, Y., Taniguchi, T., Watanabe, K. & Nagashio, K. Anisotropic Dielectric Breakdown Strength of Single Crystal Hexagonal Boron Nitride. *ACS Appl. Mater. Interfaces* **8**, 27877–27884 (2016).

22. Murrell, M. P. *et al.* Spatially resolved electrical measurements of SiO2 gate oxides using atomic force microscopy. *Appl. Phys. Lett.* **62**, 786–788 (1993).

23. Illarionov, Y. Y. *et al.* Insulators for 2D nanoelectronics: the gap to bridge. *Nature Communications* **11**, 1–15 (2020).

24. Domaretskiy, D. *et al.* Quenching the band gap of 2D semiconductors with a perpendicular electric field. 1–12 (2021).

25. Kim Le, O., Chihaia, V., On, V. V. & Ngoc Son, D. N-type and p-type molecular doping on monolayer MoS 2. *RSC Advances* **11**, 8033–8041 (2021).

26. Wang, J. *et al.* Charge Transfer within the F 4 TCNQ-MoS 2 van der Waals Interface: Toward Electrical Properties Tuning and Gas Sensing Application. *Advanced Functional Materials* **28**, 1806244 (2018).

27. Park, J. *et al.* Single-gate bandgap opening of bilayer graphene by dual molecular doping. *Advanced Materials* **24**, 407–411 (2012).

28. Wang, T. H., Zhu, Y. F. & Jiang, Q. Bandgap opening of bilayer graphene by dual doping from organic molecule and substrate. *Journal of Physical Chemistry C* **117**, 12873–12881 (2013).

29. Tian, X., Xu, J. & Wang, X. Band gap opening of bilayer graphene by F4-TCNQ molecular doping and externally applied electric field. *Journal of Physical Chemistry B* **114**, 11377–11381 (2010).

30. Zhang, W. *et al.* Opening an electrical band gap of bilayer graphene with molecular doping. *ACS Nano* **5**, 7517–7524 (2011).

31. Nourbakhsh, A., Cantoro, M., Heyns, M. M., Sels, B. F. & De Gendt, S. (Invited) Toward Ambient-Stable Molecular Gated Graphene-FET: A Donor/Acceptor Hybrid Architecture to Achieve Bandgap in Bilayer Graphene. *ECS Transactions* **53**, 121–129 (2013).

32. Ju, L. *et al.* Photoinduced doping in heterostructures of graphene and boron nitride. *Nature Nanotechnology* **9**, 348–352 (2014).

33. Roch, J. G. *et al.* Spin-polarized electrons in monolayer MoS 2. *Nature Nanotechnology* **14**, 432–436 (2019).

34. Epping, A. *et al.* Quantum transport through MoS2 constrictions defined by photodoping. *Journal of Physics Condensed Matter* **30**, 205001 (2018).

35. Guo, Y. *et al.* Charge trapping at the MoS2-SiO2 interface and its effects on the characteristics of MoS2 metal-oxide-semiconductor field effect transistors. *Applied Physics Letters* **106**, 103109 (2015).

36. Illarionov, Y. Y. *et al.* The role of charge trapping in MoS2/SiO2 and MoS2/hBN field-effect transistors. *2D Materials* **3**, 035004 (2016).

37. Park, S. *et al.* Temperature-Dependent Electronic Ground-State Charge Transfer in van der Waals Heterostructures. *Advanced Materials* **33**, 2008677 (2021).

38. Park, S. *et al.* Demonstration of the key substrate-dependent charge transfer mechanisms between monolayer MoS2 and molecular dopants. *Commun Phys* **2**, 1–8 (2019).

39. Mouri, S., Miyauchi, Y. & Matsuda, K. Tunable photoluminescence of monolayer MoS2 via chemical doping. *Nano Letters* **13**, 5944–5948 (2013).

40. Katzer, M. *et al.* Impact of dark excitons on F\"orster-type resonant energy transfer between dye molecules and atomically thin semiconductors. *Phys. Rev. B* **107**, 035304 (2023).

41. Greben, K., Arora, S., Harats, M. G. & Bolotin, K. I. Intrinsic and Extrinsic Defect-Related Excitons in TMDCs. *Nano Letters* **20**, 2544–2550 (2020).

42. Ross, J. S. *et al.* Electrical control of neutral and charged excitons in a monolayer semiconductor. *Nature Communications* **4**, 1–6 (2013).

43. Katsch, F. & Knorr, A. Excitonic theory of doping-dependent optical response in atomically thin semiconductors. *Phys. Rev. B* **105**, 045301 (2022).

39. Sponfeldner, L. *et al.* Capacitively and inductively coupled excitons in bilayer MoS$_2$. Phys. Rev. Lett. 129, 107401 (2022).

40. Hagel, J., Brem, S. & Malic, E. Electrical tuning of moir\'e excitons in MoSe$_2$ bilayers. 2D Mater. 10, 014013 (2023).

46. Katsch, F., Selig, M., Carmele, A. & Knorr, A. Theory of Exciton–Exciton Interactions in Monolayer Transition Metal Dichalcogenides. *physica status solidi (b)* **255**, 1800185 (2018).

47. Hagel, J., Brem, S., Linderälv, C., Erhart, P. & Malic, E. Exciton landscape in van der Waals heterostructures. *Phys. Rev. Research* **3**, 043217 (2021).

48. Helmrich, S. *et al.* Phonon-Assisted Intervalley Scattering Determines Ultrafast Exciton Dynamics in MoSe2 Bilayers. *Phys. Rev. Lett.* **127**, 157403 (2021).

49. Sung, J. *et al.* Broken mirror symmetry in excitonic response of reconstructed domains in twisted MoSe2/MoSe2 bilayers. *Nat. Nanotechnol.* **15**, 750–754 (2020).

50. Gutiérrez-Lezama, I., Ubrig, N., Ponomarev, E. & Morpurgo, A. F. Ionic gate spectroscopy of 2D semiconductors. *Nature Reviews Physics* **3**, 508–519 (2021).

51. Chen, D. *et al.* Excitonic insulator in a heterojunction moiré superlattice. *Nat. Phys.* **18**, 1171–1176 (2022).

52. Jia, Y. *et al.* Evidence for a monolayer excitonic insulator. *Nat. Phys.* **18**, 87–93 (2022).

53. Kohn, W. Excitonic Phases. *Phys. Rev. Lett.* **19**, 439–442 (1967).

54. Jérome, D., Rice, T. M. & Kohn, W. Excitonic Insulator. *Phys. Rev.* **158**, 462–475 (1967).

55. Castellanos-Gomez, A. *et al.* Deterministic transfer of two-dimensional materials by all-dry viscoelastic stamping. *2D Mater.* **1**, 011002 (2014).

56. Rosenberger, M. R. *et al.* Nano-"Squeegee" for the Creation of Clean 2D Material Interfaces. *ACS Appl. Mater. Interfaces* **10**, 10379–10387 (2018).

57. Extrinsic Localized Excitons in Patterned 2D Semiconductors - Yagodkin - 2022 - Advanced Functional Materials - Wiley Online Library. https://onlinelibrary.wiley.com/doi/full/10.1002/adfm.202203060.